# A Charge Domain P-8T SRAM Compute-In-Memory with Low-Cost DAC/ADC Operation for 4-bit Input Processing

Joonhyung Kim
*School of Electrical Engineering*
*Korea University*
Seoul, Korea
sjgo357@korea.ac.kr

Kyeongho Lee
*School of Electrical Engineering*
*Korea University*
Seoul, Korea
rudgh0143@korea.ac.kr

Jongsun Park
*School of Electrical Engineering*
*Korea University*
Seoul, Korea
jongsun@korea.ac.kr

## ABSTRACT

This paper presents a low cost PMOS-based 8T (P-8T) SRAM Compute-In-Memory (CIM) architecture that efficiently performs the multiply-accumulate (MAC) operations between 4-bit input activations and 8-bit weights. First, bit-line (BL) charge-sharing technique is employed to design the low-cost and reliable digital-to-analog conversion of 4-bit input activations in the proposed SRAM CIM, where the charge domain analog computing provides variation tolerant and linear MAC outputs. The 16 local arrays are also effectively exploited to implement the analog multiplication unit (AMU) that simultaneously produces 16 multiplication results between 4-bit input activations and 1-bit weights. For the hardware cost reduction of analog-to-digital converter (ADC) without sacrificing DNN accuracy, hardware aware system simulations are performed to decide the ADC bit-resolutions and the number of activated rows in the proposed CIM macro. In addition, for the ADC operation, the AMU-based reference columns are utilized for generating ADC reference voltages, with which low-cost 4-bit coarse-fine flash ADC has been designed. The 256×80 P-8T SRAM CIM macro implementation using 28nm CMOS process shows that the proposed CIM shows the accuracies of 91.46% and 66.67% with CIFAR-10 and CIFAR-100 dataset, respectively, with the energy efficiency of 50.07-TOPS/W.

## 1 INTRODUCTION

Recently, Compute-In-Memory (CIM) techniques have become one of the most promising solutions to get over the memory wall problems when designing deep neural network (DNN) accelerators. The main idea of CIM is to perform a part of or all the computations of DNN inside memories to reduce the data transfer between memories and processors. In the static random-access memory (SRAM) based CIMs [1]-[2], input activations are applied on word-lines (WLs) while the weight values are stored in SRAM bit-cells. Then, MAC operations are performed inside SRAM and the results are represented as analog voltages on bit-lines (BLs). In the typical SRAM CIM, since WLs and weight storage nodes usually represent 1-bit data, the computations between binarized inputs and weights (such as BNN) are commonly performed [1]-[2], where the DNN accuracy degradation is one of the main design bottlenecks.

To improve DNN accuracies, SRAM CIMs need to process multi-bit input-activations and weight precisions. Several previous works have proposed multi-bit SRAM CIM by utilizing high resolution DAC/ADCs [3]-[8]. However, the multi-bit SRAM CIMs [3]-[8] incur significant hardware and energy overhead since they typically need expensive digital-to-analog converter (DAC) for input activations and analog-to-digital converter (ADC) to read the analog partial MAC results. In addition, when designing SRAM CIM using multi-bit precisions, the energy efficiency and inference accuracies are mainly decided by the multiple design variables that are the bit-precision of input activations, the number of activated rows in the CIM macros, the partial sum quantization scheme, ADC bit-resolution, etc. So, for the energy efficient design of CIMs with good enough DNN accuracies, the comprehensive analysis of the multiple design variables is highly required.

In this paper, we propose the PMOS-based 8T (P-8T) SRAM Compute-In-Memory (CIM) architecture that can process 4-bit input activations and 8-bit weights with very low hardware cost. In the proposed SRAM CIM, the analog multiplication unit (AMU) that consists of the 16 local arrays, utilizes BL charge-sharing technique for low-cost digital-to-analog conversion. Using the charge domain in-SRAM operations, the AMU enables cost effective digital-to-analog conversion without employing conventional DAC circuits. The charge domain operations in AMU also improves variation tolerance, thus generating linear MAC outputs. In addition, to improve energy efficiencies without sacrificing DNN accuracy, hardware (design parameters, analog MAC variation, ADC offset) considered system simulations are per-formed to decide the number of active rows and ADC bit-resolutions in the proposed CIM. The 4-bit coarse-fine flash ADC with in-SRAM reference voltage generation is also proposed to reduce the hardware cost of analog-to-digital conversion in the proposed CIM. The hardware implementation results of the P-8T SRAM CIM using 28nm CMOS process show reliable MAC operations between 4-bit input activations and 8-bit weights with very low energy and hardware cost.

The rest of this paper is organized as follows. Section II introduces the preliminary of multi-bit CIM. The details of the proposed P-8T SRAM CIM architecture are described in Section III. Section IV presents the hardware aware system analysis. Experimental results are shown in Section V. Finally, Section V concludes this work.

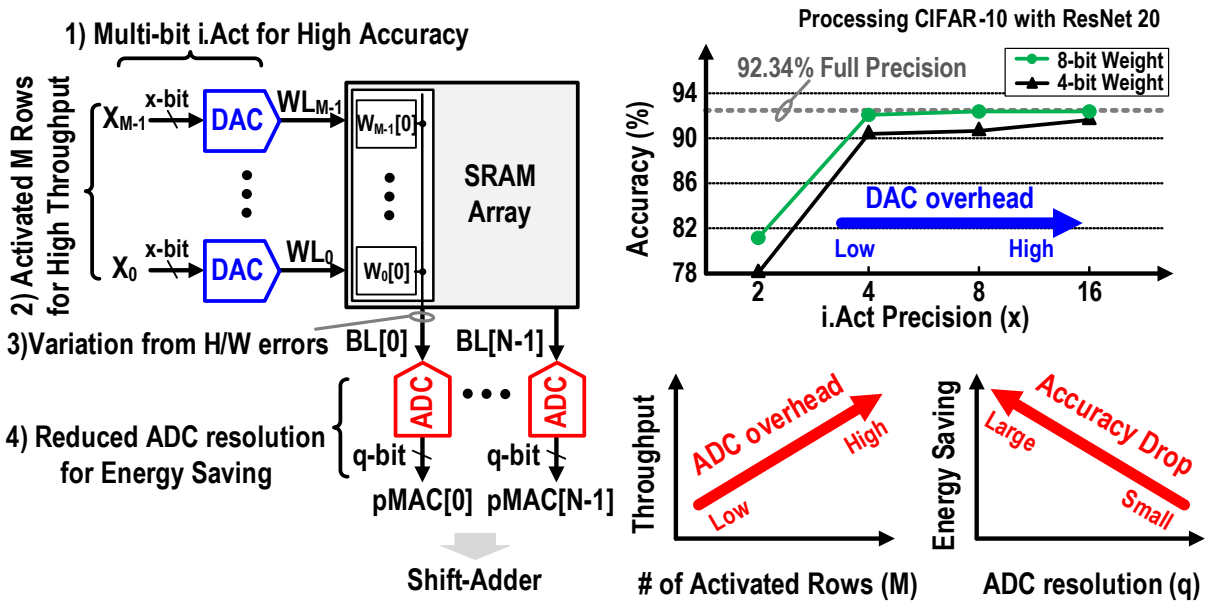

(a) (b)

**Fig. 1. (a) The typical multi-bit SRAM CIM. (b) Trade-offs between the design variables of the multi-bit SRAM CIM.**

## 2 Preliminary of SRAM Compute-In-Memory

Fig. 1.(a) shows the typical multi-bit SRAM CIM and its design considerations to improve DNN accuracies as well as energy efficiency. As shown in the figure, multi-bit input activations are first converted into multiple WL analog levels (WL voltages, pulse width) using DACs. In the SRAM array, the multi-bit weights are bit-sliced by 1-bit data and stored in the bit-cells. In order to simultaneously perform the MAC operations of different channels, the bit-sliced weight data having the same bit position is shared on the same BL. When the analog input values are applied on WLs, the MAC operations between multi-bit inputs and 1-bit weights (pMAC: partial MAC) are performed, and the results are expressed as BL analog voltage. Then, the analog voltages are converted to the digital values using ADCs. Finally, the partial MAC values are processed by the shift-adders to complete the MAC operations.

When designing the SRAM CIM that processes the high bit-precisions of inputs and weights, there exists the trade-offs between the design variables that are the bit-precision of input activations, the number of the activated rows in CIM, the partial sum quantization scheme, the ADC bit-resolution, and the hardware cost of analog circuits. As shown in the right-upper figure of Fig. 1(b), when the precision of the input activations is getting higher, the accuracy should be improved, but the area and energy cost of DAC increase as well. For example, as shown in the figure, in the case of 8-bit weight, the accuracy is similar to that of full-precision when using larger than 4-bit inputs. On the other hands, as presented in the lower figure of Fig. 1(b), the throughput of SRAM CIM is mainly decided by the number of activated rows, while the ADC cost becomes larger with increasing number of activated rows due to increasing sensing levels. In terms of the ADC bit resolution, ADC with small bit resolution saves the energy consumption of CIM at the cost of accuracy degradation. Therefore, to improve the energy efficiency and the throughput of SRAM CIM without accuracy degradation, the hardware cost of DAC and ADC should be carefully designed considering the design variables of CIM.

In the proposed SRAM CIM design, the BL charge sharing in-SRAM based digital-to-analog conversion is proposed for the low cost implementation of 4-bit DAC. In addition, with the hardware considered system simulations, the partial MAC outputs of 241 levels can be read using 4-bit ADC with negligible accuracy degradation, thus leading to dramatic ADC hardware cost reduction.

## 3 THE PROPOSED P 8T SRAM CIM ARCHITECTURE

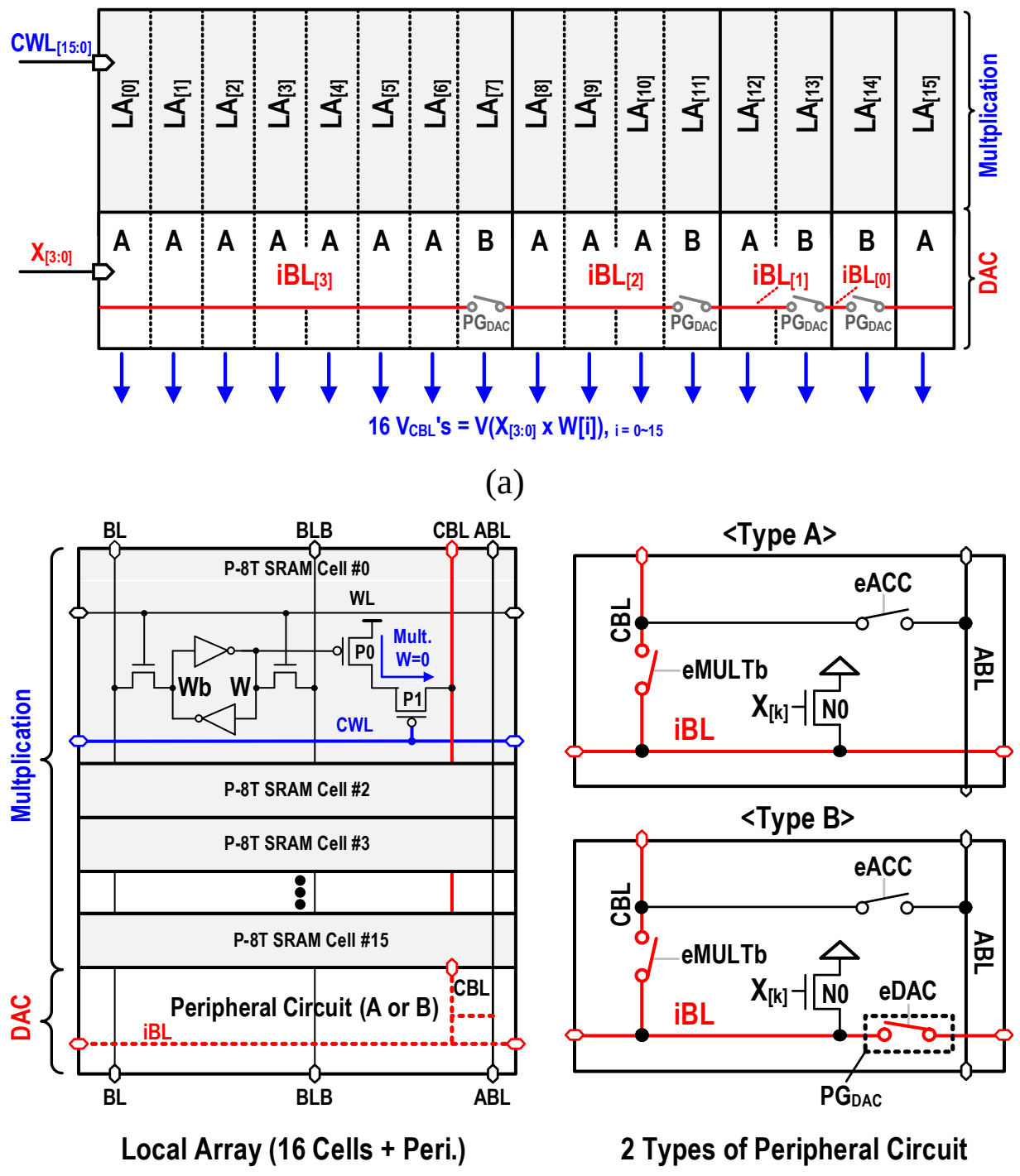

(a)

(b)

**Fig. 2. (a) The proposed analog multiplication unit (AMU) consisting of the 16 local arrays ($LA_{[0]}$ ~ $LA_{[15]}$). (b) The local array and different type of local peripheral circuits.**

### 3.1 The 4-bit DAC and its MAC Operation using BL Charge-Sharing

Fig. 2(a) presents the proposed the analog multiplication unit (AMU) consisting of the 16 local arrays ($LA_{[0]}$ ~ $LA_{[15]}$), where each array has 16 PMOS (P)-8T SRAM bit-cells and a local peripheral circuit as shown in Fig. 2(b). The proposed P-8T SRAM bit-cell includes a conventional 6T SRAM for read/write operation and two PMOS transistors ($P_0$ and $P_1$) for multiplication. In the normal SRAM mode, it operates the same as the conventional 6T SRAM array.

First, each CBL is connected with an iBL, and both BLs (all the CBL's and IBL's in Fig. 3 (a)) are initially precharged to VDD. Then, the input data ($X_{[3:0]}$ of Fig. 3(a)) is applied to the NMOS transistor of the local peripheral circuit ($N_0$ of Fig. 2(b)) to perform the digital-to-analog conversion of 4-bit input. Depending on the inputs $X_{[3:0]}$, each NMOS transistor in local peripheral circuits ($N_0$~$N_{15}$) shown in Fig. 3 (a), discharges or maintains the CBL voltage state. Please note that 8 CBL capacitances ($C_{CBL[0]}$, … , $C_{CBL[7]}$) mean MSB 1 bit ($X_{[3]}$) representing 23 value, 4 CBL capacitances ($C_{CBL[8]}$, … , $C_{CBL[11]}$) show $X_{[2]}$ representing 22 value, 2 CBL capacitances ($C_{CBL[12]}$, $C_{CBL[13]}$) mean $X_{[1]}$ showing 21 value, and an CBL capacitance ($C_{CBL[14]}$) represents LSB ($X_{[0]}$), which is illustrated in Fig. 3(a). $C_{CBL[15]}$ always maintains the precharged state. Here, the number of the charged capacitances among 16 CBL's ($C_{CBL[0]}$, … ,$C_{CBL[15]}$) represents the input activations values of $X_{[3:0]}$. For example, when the 4-bit input is '$1000_{(2)}$', the voltage levels of 8 CBL capacitances ($C_{CBL[0]}$, … , $C_{CBL[7]}$) are discharged, and the rest of the capacitances ($C_{CBL[8]}$, … ,$C_{CBL[15]}$) maintain the precharged state. The capacitances representing different digits are separated using the switch denoted as $PG_{DAC}$ of Type B peripheral circuit shown in Fig. 2 (b). In other words, the local peripheral circuit type B disconnects between two different iBLs. For example, as shown in Fig. 2(a), the $LA_{[7]}$ (between $iBL_{[3]}$ and $iBL_{[2]}$) has a type B peripheral circuit, which is same as $LA_{[1]}$, $LA_{[3]}$, and $LA_{[4]}$, while the rests have a type A peripheral circuit.

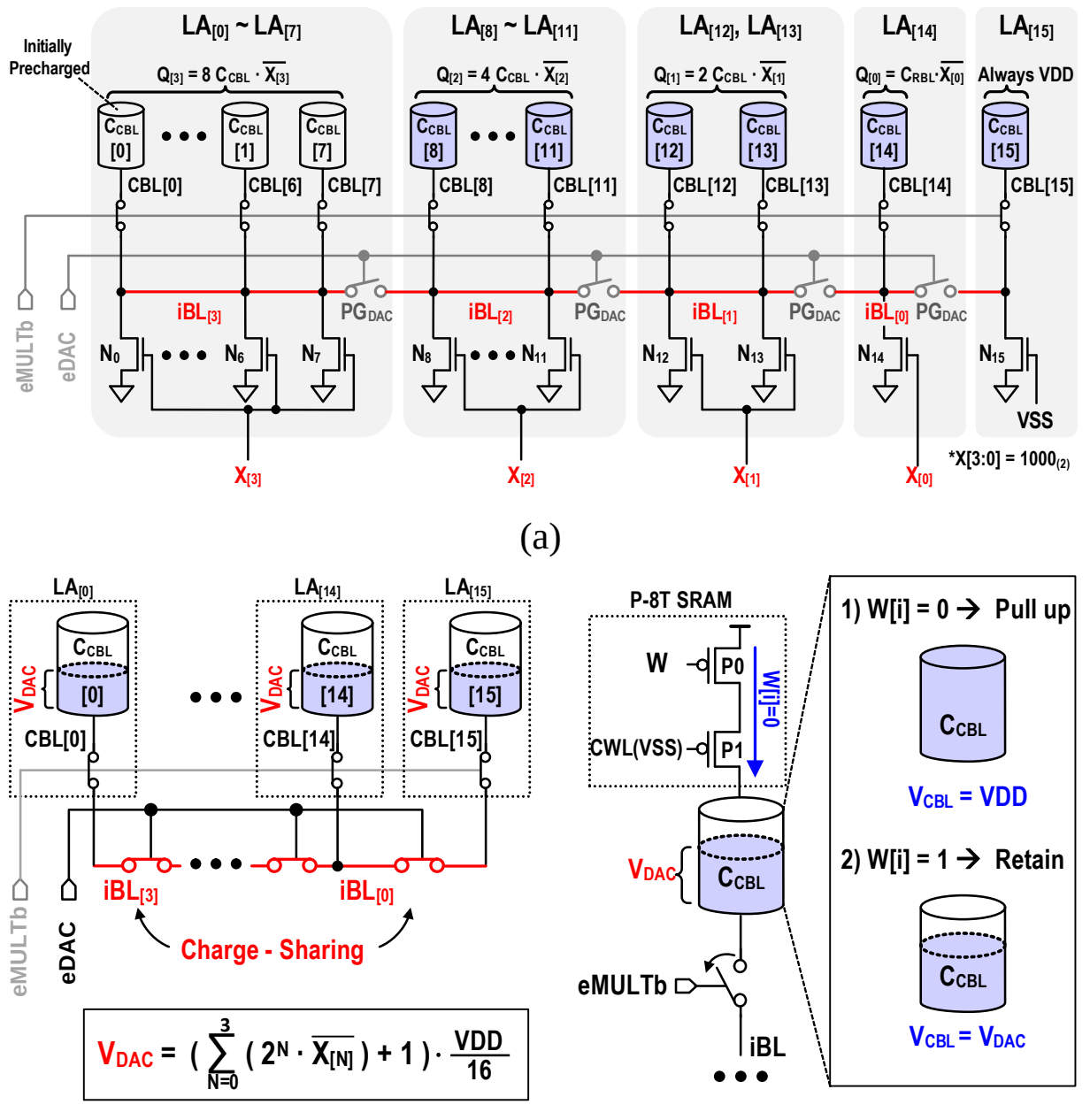

(a)

(b) (c)

**Fig. 3. The multiplication operation using AMU: (a) Input evaluation, (b) Digital-to-analog conversion of 4-bit input, (c) Multiplication.**

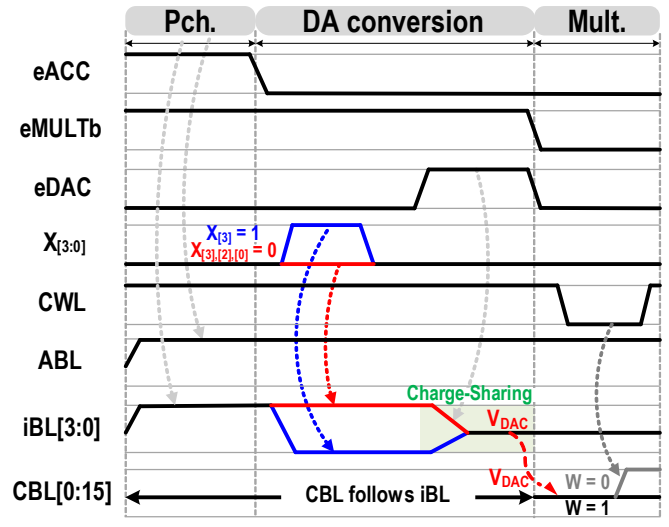

| 4-bit i.Act $X_{[3:0]}$ | 1-bit Weight (W) | 16 level Mult. Result $V_{CBL}$ |
|---|---|---|
| don't care | 0 $V_{DAC} \rightarrow VDD$ | VDD : $0_{(10)}$ |
| 0000 | 1 $V_{DAC} \rightarrow V_{DAC}$ | VDD : $0_{(10)}$ |
| 0001 | | 15/16·VDD : $1_{(10)}$ |
| 0010 | | 14/16·VDD : $2_{(10)}$ |
| ⋮ | | ⋮ |
| 1110 | | 2/16·VDD : $14_{(10)}$ |
| 1111 | | 1/16·VDD : $15_{(10)}$ |

**Fig. 4. The operation waveform and the truth table of the multiplication.**

After the charged or discharged capacitances are decided depending on the input activation value, when the eDAC signal is 'High', the switch in the type B peripheral circuit connects different iBLs. Then, as shown in Fig. 3(b), the charge sharing is performed between all CBL capacitances, and every iBLs and CBLs in the AMU have the same voltage level representing the analog level of 4-bit input data. If the 4-bit input data is '$1000_{(2)}$', the CBL voltage level is decided as half-VDD. Using BL charge sharing in the proposed AMU, the analog voltage levels of 4-bit input can be expressed as: $V_{DAC} = (\sum_{i=0}^{3}(2^i \times \overline{X[i]}) + 1) \times \frac{VDD}{16}$.

In the proposed AMU, the multiplication between the 4-bit input activation and 1-bit weight is performed through P-8T bitcell, and the result is represented as the CBL analog voltage. The detailed process of the multiplication is illustrated in Fig. 3(c). When the multiplication process starts, eMULTb signal becomes 'Low' to disconnect iBL with CBL. Then, the CWL signal becomes 'Low' to activate $P_1$ of Fig. 3(c). When the weight value is '1', the previous CBL voltage (4-bit input value) is preserved since the $P_0$ transistor is turned off. When the weight value is '0', the $P_0$ transistor is turned on, and CBL capacitance is filled with VDD representing zero multiplication result. So, the multiplication result between the 4-bit input activation and 1-bit weight is represented by 16 different analog voltages of the CBL. As shown in Fig. 2 (b), the proposed AMU simultaneously performs

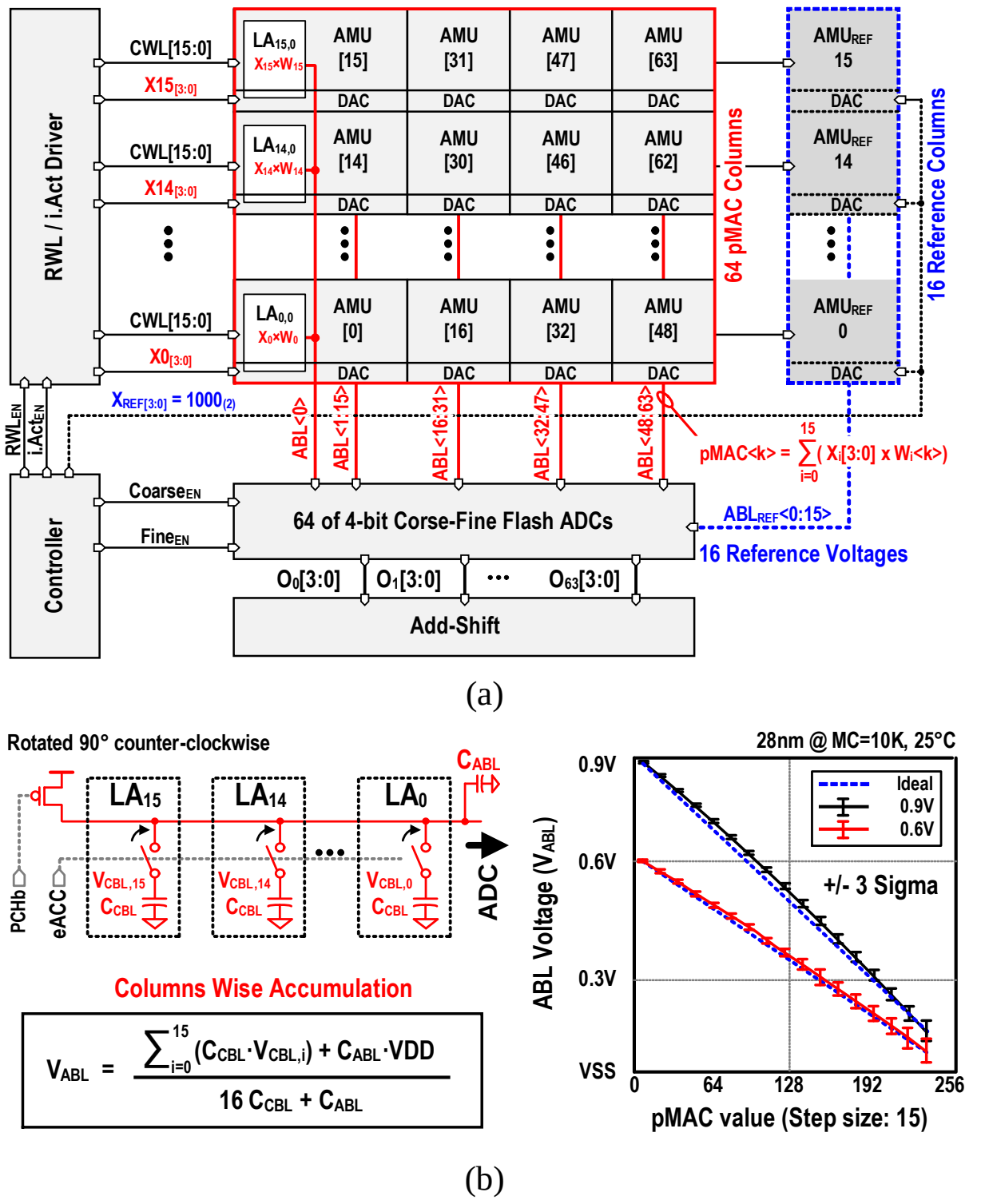

(a)

(b)

**Fig. 5. (a) The proposed SRAM CIM architecture. (b) The circuit model and its equation of the accumulation operation, and the comparison between them.**

16 multiplications. The operating waveform and the truth table of the overall multiplication operation is summarized in Fig. 4. As shown in the figure, the multiplication using the AMU can be divided into the precharge phase (Pch.), digital-to-analog conversion phase (DA conversion), and multiplication phase (Mult.). Based on BL charge-sharing operation, the AMU performs the digital-to-analog conversion of a 4-bit input data without employing the conventional DAC circuits.

Fig. 5 presents the overall architecture of the 256x80 P-8T SRAM CIM macro, the circuit model, and the equation of the accumulation operation. As shown in Fig. 5 (a), the proposed SRAM array that consists of multiple AMUs can be divided into AMU arrays for analog computation (indicated in bold) and $AMU_{REF}$ arrays for ADC reference generation (indicated in dot). The ADC reference generation using $AMU_{REF}$ arrays will be discussed in section III. B. In the proposed SRAM CIM macro, 16 different local arrays are connected to an accumulation bit-line (ABL), so the column-wise summation is performed to complete the accumulation operation. For example, $LA_{0,0}$, $LA_{1,0}$, … , $LA_{15,0}$ are connected to $ABL_0$, as shown in Fig. 5 (a). During the CIM operation, 16 4-bit input data of different channels in a layer are applied to the AMU. After each local array in AMU completes the multiplications that are stored in CBL capacitances ($C_{CBL}$'s) as depicted in Fig. 3(a), all the CBL charges in the same ABL are shared using the eACC signal. As a result, the summation of 16 4-bit multiplication results is represented as the ABL analog voltage. In the proposed SRAM CIM macro, since the partial MAC (pMAC) output has 241 levels, the 8-bit resolution ADC is needed for accurate analog readout. However, considering hardware considered system simulations, 4-bit flash ADC instead of the 8-bit ADC is adopted with ignorable accuracy loss in our design, which will be addressed in section IV. Finally, after the ADC operation and add-shift operation, the proposed SRAM CIM macro produces 8 partial sum (128 MAC operation) results between 4-bit inputs and 8-bit weights.

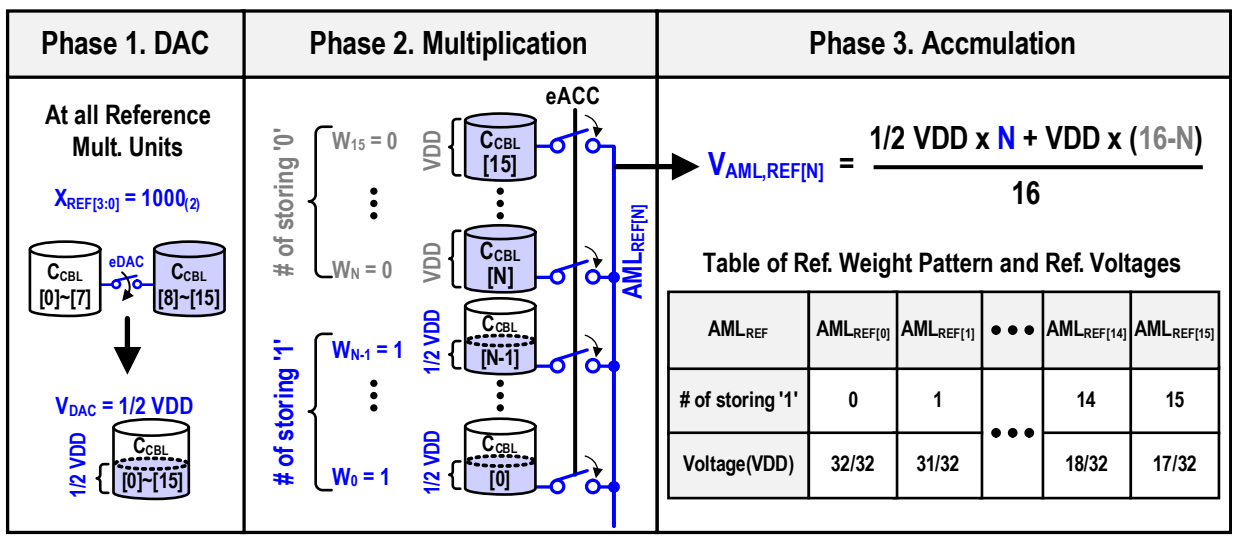

(a)

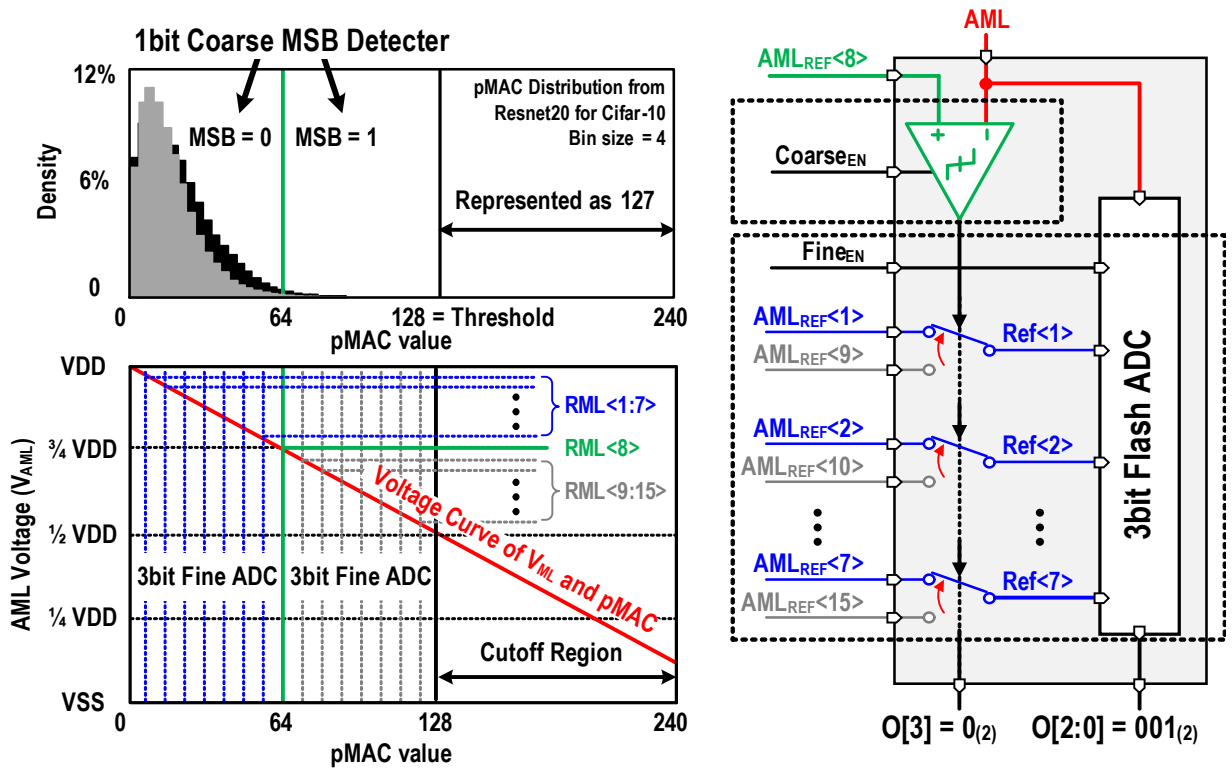

(b)

**Fig. 6. (a) The operation principle of AMU$_{REF}$ arrays for ADC reference voltage generation. (b) The proposed 4-bit coarse-fine flash ADC with the quantization method using threshold.**

Fig. 5(b) shows the comparisons between the accumulation process equation obtained from circuit model and the simulation results from Monte-Carlo simulations. As shown in the left figure of Fig. 5 (b), when the eACC signal connects CBL and ABL, all the charges in each CBL capacitance are summed onto ABL to generate the pMAC results. Based on the BL charge sharing operation, the total charge is equal to the charge-sum of 16 multiplication results stored in each CBL capacitance ($C_{CBL}$) and the charge of the $C_{ABL}$ that is precharged by VDD. The total capacitance is equal to the sum of 16 CBL capacitances and the ABL capacitance ($C_{ABL}$). Thus, $V_{ABL}$, which represents the partial MAC value, can be expressed as the equation shown in left-bottom figure of Fig. 5 (b). In order to evaluate the circuit operations of the proposed SRAM CIM macro, the comparison between 2 different voltage curves of pMAC value is shown in the right figure of Fig. 5 (b). The voltage curves are obtained from the ideal equation (in dot line) and the actual circuit simulations using 10K-samples Monte-Carlo simulation (in bold line), respectively. As shown in the figure, the voltage curve from Monte-Carlo simulations shows almost same results compared with the ideal equation curve. By utilizing BL charge sharing for accumulation together with digital-to-analog conversion and multiplication, the proposed SRAM CIM macro provide variation tolerant and linear pMAC outputs.

## 3.2 Bit-Line Charge-Sharing based ADC operation

In the proposed SRAM CIM macro, for low cost implementation of analog-to-digital conversion, we propose the 4-bit coarse-fine flash ADC based on in-SRAM reference voltage generation scheme. For the 4-bit ADC operation, 16 AMU$_{REF}$ arrays presented in Fig. 4(a) are utilized to generate 16 different ADC reference voltages. While 64 AMU arrays process the MAC operations, AMU$_{REF}$ arrays also perform almost the same operations using a different input patterns using the NMOS transistors ($N_0$~$N_{15}$ depicted in Fig. 3(a)) to generate the ADC reference voltages.

Fig. 6(a) illustrates the operation principle of AMU$_{REF}$ arrays for ADC reference voltage generation. As shown in the figure, during digital-to-analog conversion phase, the input pattern of '$1000_{(2)}$' is applied to all the AMU$_{REF}$ arrays, and the half-VDD state is formed in all RBLs ($RBL_{[0]}$ ~ $RBL_{[15]}$). In the multiplication phase, the stored data pattern of the local arrays in the same RBL determines whether the CBL voltage is 'half-$V_{DD}$' or '$V_{DD}$'. After BL charge sharing is performed, the ADC reference voltage is generated on the CBL depending on the stored data pattern. Based on charge sharing operation, the reference voltage can be expressed as equation: $V_{AML[N]} = (\frac{1}{2}N + (16-N)) \times \frac{VDD}{16}$. As shown in the table in Phase 3, since the number of storing '1' data differs by each $ABL_{REF[N]}$, 16 different voltage levels ($V_{DD}$, $^{31}/_{32}V_{DD}$, … $^{17}/_{32}V_{DD}$) are formed and used for 4-bit ADC operation.

Fig. 6(b) also shows the proposed 4-bit ADC structure that performs 1-bit coarse and 3-bit fine analog readout. Although the partial MAC (pMAC) output of the proposed CIM macro has 241 levels, most of the pMAC results are places within the range from 0 to 128 (129 levels), as shown in the left-upper figure of Fig. 6(b). Here, the pMAC distributions are obtained from 2nd layer of ResNet-20 with CIFAR-10 dataset when using the proposed SRAM CIM. To reduce the ADC bit resolution, the pMAC values that are larger than the pre-decided threshold (rarely observed) are quantized as the threshold value. The accuracy results with different threshold values are discussed in section IV. For the ADC operation, first of all, the 1-bit coarse analog readout is performed by comparing the ABL voltage having a pMAC value with $ABL_{REF[8]}$ having $^{48}/_{64}$ $V_{DD}$ to decide the MSB of pMAC. Then, based on the MSB data, we can determine whether the next 3-bit flash ADC sampling applies to the lower pMAC range (from 0 to 63) or the upper pMAC range (from 64 to 127). When the MSB is '0', the next 3-bit flash ADC sampling occurs on the lower pMAC range (from 0 to 63) using $ABL_{REF[1]}$ ~ $ABL_{REF[7]}$. Otherwise, it is on the upper pMAC range (from 64 to 127) using $ABL_{REF[9]}$ ~ $ABL_{REF[15]}$. In the proposed 4-bit ADC, only 8 comparators are needed to complete the operation.

# 4 HARDWARE AWARE SYSTEM ANALYSIS

As mentioned in Section II, the energy efficiency and inference accuracies of SRAM CIM are mainly affected by ADC bit-resolution and its hardware cost. In the proposed SRAM CIM, the partial sum quantization method and the ADC bit-resolution are investigated based on the hardware considered system simulations that take into account the number of activated rows in CIM macro and the hardware errors (PVT error and comparator offset). Here, the hardware errors are obtained from the 10K Monte-Carlo simulation shown in Fig. 5(b). Fig. 7 presents the results of the hardware considered PyTorch simulations using ResNet-20 with CIFAR-10 dataset.

As shown in the upper-left figure of Fig. 7(a), since most of the pMAC values have small amplitude, as mentioned in Section III.B, the pMAC values that are larger than the pre-decided threshold, are quantized to the threshold value. Here, we define $cutoff = 1 - threshold/2^q$, where the q means the ADC resolution for precise pMAC readout. For example, if the cutoff value is 0.375 in the case of 4 activated rows (Th. = 24), pMAC values larger than 24 are clipped and represented as 23, as shown in left figure of Fig. 7(a). To find the proper cutoff value, the design variables such as the number of activated rows and hardware errors (PVT error and comparator offset) in CIM, are also considered when analyzing accuracies. In the center figure of Fig.

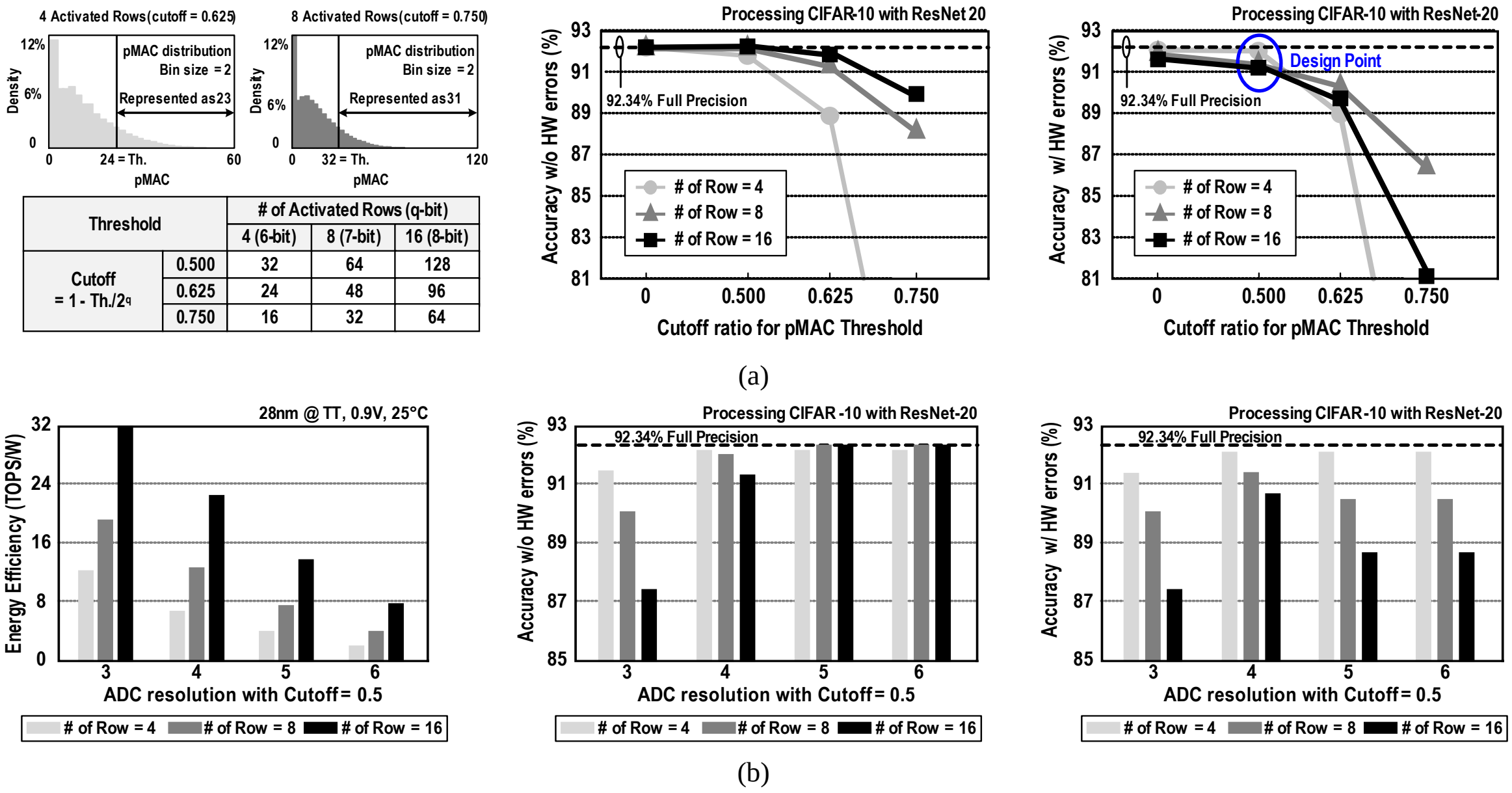

**Fig. 7. (a) (left figure) The cutoff region with the number of activated rows. (center figure) The CIFAR-10 accuracies with different cutoff values when considering only the number of activated rows. (right figure) The CIFAR-10 accuracies with different cutoff values when considering both of the number of activated rows and hardware errors. (b) (left) The energy efficiencies with different ADC bit resolutions and the number of activated rows using the cutoff of 0.50. (center) The accuracies with different ADC bit resolutions and the number of activated rows using the cutoff of 0.50. (right) The accuracies with different ADC bit resolutions and the number of activated rows when considering the hardware errors using the cutoff of 0.50.**

| | |
|---|---|
| Technology | 28nm CMOS |
| Supply Voltage | 0.6V - 1.2V |
| SRAM Array Size | 256×80 (16×5 AMUs) |
| Area | 0.0324mm$^2$ |
| Input Precision | 4-bit |
| Weight Precision | 8-bit |
| # of Activated Rows | 8, 16 |
| DAC Scheme | Charge-sharing |
| MAC Scheme | Charge-sharing |
| ADC Scheme | 4-bit Coarse-Fine Flash |
| Energy Efficiency (16 Rows) | 50.07 (0.6V) - 9.77 (1.2V) |

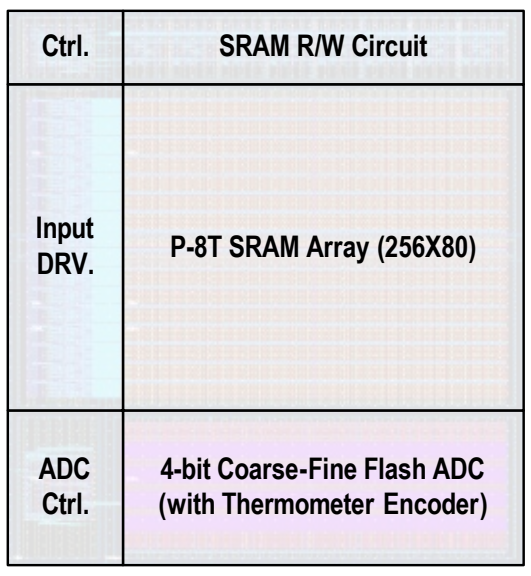

**Fig. 8. The implementation of the proposed CIM and the layout configuration.**

7(a), the accuracy analysis is performed only considering the number of activated rows and varying cutoff values. When the cutoff value is 0.625, the accuracy decreases smaller than 1% regardless of the number of the activated rows. However, when hardware errors are considered (the right figure of Fig. 7 (a)), at the cutoff value of 0.50, the accuracy degrades around 1%. So, in our design, the partial sum quantization with the cutoff value of 0.50 is used.

Fig. 7(b) shows the energy efficiency (left figure) and the accuracy (center and right figures) results for different ADC bit-resolutions and the different number of activated rows at the cutoff value of 0.50. As shown in the figures, without considering the hardware errors (the center figure of Fig. 7(b)), regardless of the number of the activated rows, as ADC bit resolution becomes larger, the accuracies approach to the baseline (full precision) while degrading the energy efficiencies. However, when the hardware errors are considered, even if the ADC resolutions increases, accuracy drop is observed with increasing number of activated rows. In addition, as shown in the right figure of Fig. 7(b), when the number of the activated rows is 4, the accuracy using 4-bit ADC is similar to the accuracies obtained using 5-bit or 6-bit ADCs. It is because the higher bit resolution of ADC makes the sensing margin smaller, resulting it more susceptibility to hardware errors. So, in the case of the number of the activated rows being 8 and 16, the best accuracy is observed when 4-bit ADC is used. Based on the hardware considered system simulations, the proposed SRAM CIM uses the cutoff value of 0.5 with 4-bit ADC resolution.

# 5 EXPERIMENTAL RESULTS

The 256×80 P-8T SRAM CIM macro has been implemented using 28nm CMOS process. Fig. 8 shows the layout and the summary of the implementation. The proposed SRAM CIM utilizes the charge domain operations for the 4-bit input digital-to-analog conversions and for the MAC operations between the inputs and 8-bit weights. The inference accuracies of the proposed SRAM CIM using CIFAR-10 and CIFAR-100 data with ResNet-20, are shown in the Table I. In the case of activating 8 rows, the accuracies of CIFAR-10 and CIFAR-100 are 91.46% and 66.67%, respectively. When activating 16 rows for high throughput, the accuracies of CIFAR-10 and CIFAR-100 are 90.47% and 64.26%, respectively.

Fig. 9(a) presents the reliability analysis of the digital-to-analog conversion using AMU when performing 10K-samples Monte-Carlo simulations. Thanks to the BL charge-sharing operation, the proposed digital-to-analog conversion is seamlessly performed at different supply voltages. The worst standard deviation of 1.8mV is observed when the DAC code is 8 and 0.6V, leading to negligible accuracy degradation. Fig. 9(b) also shows the comparison of the energy consumption between different 4-bit flash ADCs. The proposed ADC effectively reduces the energy consumption of the reference voltage generation using in-SRAM reference generation. It also reduces the number of comparators by performing the 1-bit coarse sampling and 3-bit fine sampling. Compared with the conventional 4-bit flash ADC using R-ladder, the proposed coarse-fine (C-F) ADC saves a 43.9% of the normalized energy consumption.

The energy efficiencies and the operating frequencies of the proposed SRAM CIM when running with different supply voltages are presented in in Fig. 10(a). In the case of 16 activated rows, the energy efficiencies of 9.77-TOPS/W and 50.07-TOPS/W are achieved with 1.2V and 0.6V supply voltages, respectively. The operating frequency ranges from 76.9MHz at 0.6V and 435MHz at 1.2V. As shown in Fig. 10(b), AMU with 256×80 array that performs digital-to-analog conversions and MAC operations, consumes only 11.4% of the total energy. The ADC delay accounts for only 31.8% of the total delay. The comparisons to prior state-of-the-art multi-bit SRAM CIMs [4], [5] are also shown in Table II. Based on the charge domain DAC and MAC operations, our work can perform the CIM operations in wide range of supply voltages (0.6V ~1.2V). By employing 4-bit coarse-fine flash ADC that obtains the lowest ADC resolution compared to other works, the proposed P-8T SRAM CIM achieves 22.19-TOPS/W of energy efficiency with 4.4ns of cycle time.

TABLE I
**THE CIFAR-10 AND CIFAR-100 ACCURACIES WITH RESNET-20 WHEN THE PROPOSED SRAM CIM IS USED.**

| Resnet-20 (Baseline Accuracy) | | Cifar-10 (92.34%) | | Cifar-100 (68.81%) | |
|---|---|---|---|---|---|
| # of Activated Rows | | 8 | 16 | 8 | 16 |
| Inference Accuracy w/ Cufoff = 0.5 | w/o HW errors | 92.01% | 91.06% | 67.55% | 65.26% |
| | w/ HW errors | 91.46% | 90.47% | 66.67% | 64.26% |

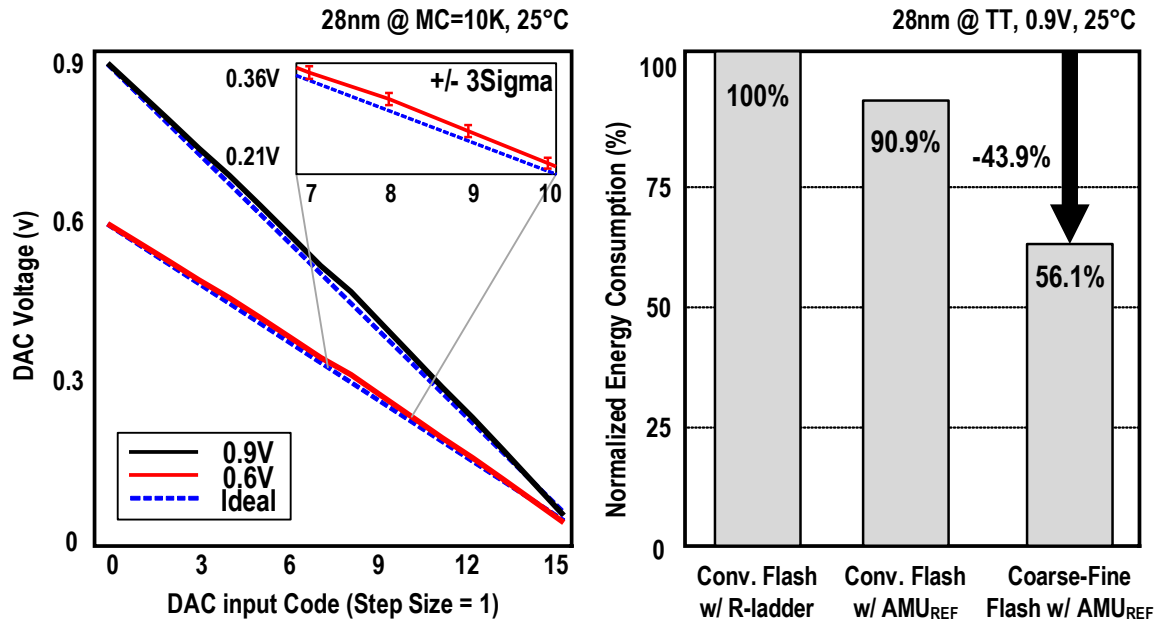

(a) (b)

**Fig. 9. (a) The reliability analysis of the DAC operation using AMU. (b) The comparison of the energy consumption between different 4-bit flash ADCs.**

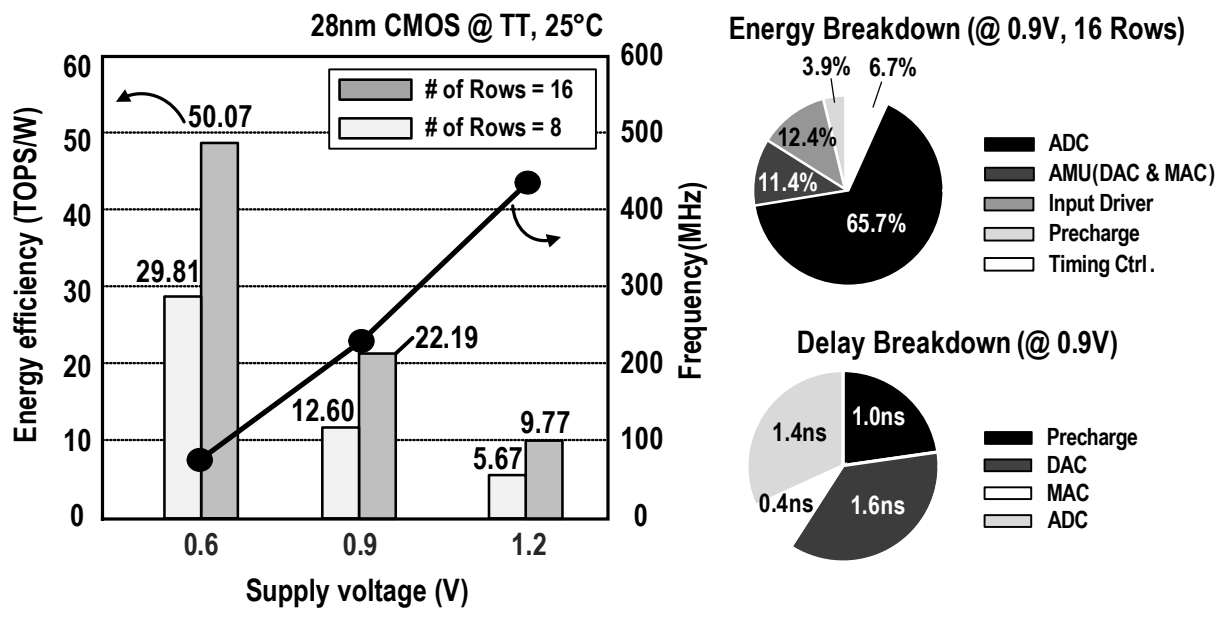

(a) (b)

**Fig. 10. (a) The energy efficiencies and the operating frequencies of the proposed SRAM CIM at different supply voltages. (b) The energy consumption breakdown and the delay breakdown of the proposed SRAM CIM.**

# 6 CONCLUSIONS

In this paper, we present PMOS based 8T SRAM CIM architecture that efficiently performs the multiply-accumulate (MAC) operations between 4-bit input activations and 8-bit weights. In the proposed CIM macro, based on the BL charge-sharing operation, the analog multiplication unit (AMU) performs the MAC operations, and it also generates the ADC reference voltages. The BL charge sharing based digital-to-analog conversion can represent 4-bit inputs as BL analog voltages, alleviating the DAC implementation cost. In addition, the 4-bit coarse-fine flash ADC using in-SRAM ADC reference voltages can convert partial MAC outputs of 241 levels into 4-bit digital outputs, which reduce the ADC cost while maintaining inference accuracies. Hardware-considered system simulations is also employed to decide the CIM design parameters that leads to highly energy efficient CIM design without sacrificing DNN accuracies.

TABLE II
**COMPARISON TO PRIOR STATE-OF-THE-ART MULTI-BIT SRAM CIMS**

| Reference | | J. Su et al. [4] | Z. Chen et al. [5] | This work |
|---|---|---|---|---|
| Technology | | 28nm | 65nm | 28nm |
| Cell Type | | 6T + Two-way Transpose | 6T + Local peri. | P-8T + Local peri. |
| Input / Weight | | 4 bit / 8 bit | 4 bit / 8 bit | 4 bit / 8 bit |
| Computing Scheme | DAC | Multi level Source | Current Steering | Charge-sharing |
| | MAC | Cureent | Charge-sharing | Charge-sharing |
| Supply Voltage | | 0.85V, 0.95V | 1.2V | 0.6-1.2V |
| ADC | | 5-bit SAR ADC | 6-bit Ci-SAR ADC | 4-bit C-F Flash ADC |
| Cycle time | | 8.6ns (0.85V) | 14.3n (1.2V) | 4.4ns (0.9V) |
| Array Size | | 8KB(512×128) | 8KB(512×128) | 4.5KB (256×80) |
| # of Activated Channels | | 16 Rows | 128 Columns | 8, 16 Rows |
| GOPS/2KB | | 29.8 (0.85V) | 143.35 (1.2V) | 45.54 (0.9V) 89.04 (1.2V) |
| TOPS/W | | 15.17 (0.85V) 14 (0.95V) | 6.18 (1.2V) | 50.07 (0.6V) 22.19 (0.9V) |
| Accuracy[1)] | | 91.5 – 91.9 % | 89.0 % | 90.47 – 91.46 % |

*1) CIFAR-10 on ResNet-20 with hardware error*

# ACKNOWLEDGEMENTS

This research was supported by the National Research Foundation of Korea grant funded by the Korea government (No. NRF-2020R1A2C3014820. The EDA tool was supported by the IC Design Education Center (IDEC), Korea.

# REFERENCES

[1] Z. Jiang, S. Yin, J.-S. Seo, and M. Seok, ‘C3SRAM: An In-Memory-Computing SRAM Macro Based on Robust Capacitive Coupling Computing Mechanism’, IEEE Journal of Solid-State Circuits, vol. 55, no. 7, pp. 1888–1897, Jul. 2020.

[2] S. Yin, Z. Jiang, J.-S. Seo, and M. Seok, ‘XNOR-SRAM: In-Memory Computing SRAM Macro for Binary/Ternary Deep Neural Networks’, IEEE Journal of Solid-State Circuits, vol. 55, no. 6, pp. 1733–1743, Jun. 2020.

[3] A. Biswas and A. P. Chandrakasan, ‘CONV-SRAM: An Energy-Efficient SRAM With In-Memory Dot-Product Computation for Low-Power Convolutional Neural Networks’, IEEE Journal of Solid-State Circuits, vol. 54, no. 1.

[4] J.-W. Su et al., ‘Two-Way Transpose Multibit 6T SRAM Computing-in-Memory Macro for Inference-Training AI Edge Chips’, IEEE Journal of Solid-State Circuits, vol. 57, no. 2, pp. 609–624, Feb. 2022.

[5] Z. Chen et al., ‘CAP-RAM: A Charge-Domain In-Memory Computing 6T-SRAM for Accurate and Precision-Programmable CNN Inference’, IEEE Journal of Solid-State Circuits, vol. 56, no. 6, pp. 1924–1935, Jun. 2021.

[6] X. Si et al., ‘A Local Computing Cell and 6T SRAM-Based Computing-in-Memory Macro With 8-b MAC Operation for Edge AI Chips’, IEEE Journal of Solid-State Circuits, vol. 56, no. 9, pp. 2817–2831, Sep. 2021.

[7] J.-W. Su et al., ‘16.3 A 28nm 384kb 6T-SRAM Computation-in-Memory Macro with 8b Precision for AI Edge Chips’, in 2021 IEEE International Solid- State Circuits Conference (ISSCC), Feb. 2021, vol. 64, pp. 250–252.

[8] J. Lee, H. Valavi, Y. Tang, and N. Verma, ‘Fully Row/Column-Parallel In-memory Computing SRAM Macro employing Capacitor-based Mixed-signal Computation with 5-b Inputs’, in 2021 Symposium on VLSI Circuits, Jun. 2021, pp. 1–2.